%% file: main.tex
\documentclass[9pt,conference]{IEEEtran}
\IEEEoverridecommandlockouts
\usepackage{cite}

\usepackage{amsmath,amssymb,amsfonts}

\usepackage{graphicx}
\usepackage{mathabx}
\usepackage{textcomp}
\usepackage{balance}
\usepackage{xcolor}
\usepackage{algorithm}
\usepackage[noend]{algpseudocode}

\usepackage{mathtools}
\usepackage{multirow}
\usepackage{subcaption}
\usepackage{calc}
\usepackage{comment}
\usepackage[colorinlistoftodos]{todonotes}
\usepackage{adjustbox}
\usepackage[bookmarks=false]{hyperref}

\usepackage{float}
\usepackage{wrapfig}
\hypersetup{
    colorlinks = true,
}

\newsavebox\MBox

\usepackage{MnSymbol,pifont}

\usepackage{xurl}

\input{comments.tex}

\usepackage{etoolbox}
\makeatletter
\patchcmd{\@makecaption}
  {\\}
  {.\ }
  {}
  {}
\makeatother

\usepackage{listings}
\usepackage{multicol}
\usepackage{nicematrix}
\usepackage{array, booktabs, tabularx,makecell}
\usepackage{enumitem}

\setenumerate[1]{itemsep=0.5pt,partopsep=0pt,parsep=\parskip, topsep=2pt, leftmargin=12pt,}

\usepackage{listings}
\definecolor{dkgreen}{rgb}{0,0.6,0}
\definecolor{gray}{rgb}{0.5,0.5,0.5}
\definecolor{mauve}{rgb}{0.58,0,0.82}

\lstdefinestyle{interfaces}{
  float=tp,
  floatplacement=tbp,
}

\usepackage{siunitx}

\makeatletter
\def\BState{\State\hskip-\ALG@thistlm}
\makeatother
\definecolor{codegreen}{rgb}{0,0.6,0}
\definecolor{codegray}{rgb}{0.5,0.5,0.5}
\definecolor{codepurple}{rgb}{0.58,0,0.82}
\definecolor{backcolour}{rgb}{0.95,0.95,0.92}
\definecolor{backcolour}{rgb}{0.95,0.95,0.92}

\usepackage{xcolor}
\definecolor{commentgreen}{RGB}{2,112,10}
\definecolor{eminence}{RGB}{108,48,130}
\definecolor{weborange}{RGB}{255,165,0}
\definecolor{frenchplum}{RGB}{129,20,83}
\definecolor{bostonuniversityred}{rgb}{0.8, 0.0, 0.0}
\usepackage{listings}
\usepackage{textcomp}
\lstdefinestyle{mystyle}{
    backgroundcolor=\color{backcolour},   
    commentstyle=\color{codegreen},
    keywordstyle=\color{magenta},
    numberstyle=\tiny\color{codegray},
    stringstyle=\color{codepurple},
    basicstyle=\ttfamily\footnotesize,
    breakatwhitespace=false,         
    breaklines=true,                 
    captionpos=b,                    
    keepspaces=true,                 
    numbers=left,                    
    numbersep=5pt,                  
    showspaces=false,                
    showstringspaces=false,
    showtabs=false,                  
    tabsize=2,
    captionpos=t,
    moredelim=**[is][\color{red}]{@}{@}1
}
\begin{document}

%
\title{\LARGE
AIM: Accelerating \underline{A}rbitrary-precision \underline{I}nteger \underline{M}ultiplication on \\ Heterogeneous Reconfigurable Computing Platform Versal ACAP
}


\author{\IEEEauthorblockN{Zhuoping Yang\IEEEauthorrefmark{1}, Jinming Zhuang\IEEEauthorrefmark{1},
Jiaqi Yin\IEEEauthorrefmark{2}, Cunxi Yu\IEEEauthorrefmark{2}, 
Alex K. Jones\IEEEauthorrefmark{1} and Peipei Zhou\IEEEauthorrefmark{1}}
\IEEEauthorblockA{\IEEEauthorrefmark{2}University of Maryland, College Park, USA,
\{jyin629,cunxiyu\}@umd.edu}
\IEEEauthorblockA{\IEEEauthorrefmark{1}University of Pittsburgh, USA,
\{zhuoping.yang, jinming.zhuang, akjones, peipei.zhou\}@pitt.edu}
}


\maketitle


\input{0_abstract.tex}


\input{1_intro.tex}

\input{2_related}
\input{3_versal_architecture}

\input{4_AIM_single_kernel}
\input{5_AIM_framework_application}

\input{6_Experiments}
\input{7_conclusion}

\bibliographystyle{IEEEtran}
\bibliography{reference}
\end{document}

%% file: comments.tex
\usepackage{ifthen}
\usepackage{xcolor}
\usepackage{mathabx}
\newboolean{showcomments}
\setboolean{showcomments}{true}

\makeatletter
\newcommand{\mynote}[3]{%
  \ifthenelse{\boolean{showcomments}}{%
   \fbox{\bfseries\sffamily\scriptsize#1}%
   {\small$\blacktriangleright$\textsf{\emph{\color{#3}{#2}}}$\blacktriangleleft$}}%
  {%
   \@bsphack
   \@esphack
  }%
}
\makeatother

\definecolor{asparagus}{rgb}{0.53, 0.66, 0.42}

%% file: 0_abstract.tex
\begin{abstract}
Arbitrary-precision integer multiplication is the core kernel of many applications including scientific computing, cryptographic algorithms, etc. 
Existing acceleration of arbitrary-precision integer multiplication includes CPUs, GPUs, FPGAs, and ASICs. 
To leverage the hardware intrinsics low-bit function units (32/64-bit), arbitrary-precision integer multiplication can be calculated using Karatsuba decomposition, and Schoolbook decomposition by decomposing the two large operands into several small operands, generating a set of low-bit multiplications that can be processed either in a spatial or sequential manner on the low-bit function units, e.g., CPU vector instructions, GPU CUDA cores, FPGA digital signal processing (DSP) blocks. 
Among these accelerators, reconfigurable computing, e.g., FPGA accelerators are promised to provide both good energy efficiency and flexibility. 
We implement the state-of-the-art (SOTA) FPGA accelerator and compare it with the SOTA libraries on CPUs and GPUs. 
Surprisingly, 
in terms of energy efficiency, we find that the FPGA has the \emph{lowest energy efficiency}, i.e., 0.29x of the CPU and 0.17x of the GPU with the same generation fabrication. Therefore, key questions arise: \emph{Where do the energy efficiency gains of CPUs and GPUs come from? Can reconfigurable computing do better? If can, how to achieve that?}

We first identify that the biggest energy efficiency gains of the CPUs and GPUs come from the dedicated vector units, i.e., vector instruction units in CPUs and CUDA cores in GPUs. 
FPGA uses DSPs and lookup tables (LUTs) to compose the needed computation, which incurs overhead when compared to using vector units directly.
New reconfigurable computing, e.g., ``FPGA+vector units'' is a novel and feasible solution to improve energy efficiency. 
In this paper, we propose to map arbitrary-precision integer multiplication onto such a ``FPGA+vector units'' platform, i.e., AMD/Xilinx Versal ACAP architecture, a heterogeneous reconfigurable computing platform that features 400 AI engine tensor cores (AIE) running at 1 GHz, FPGA programmable logic (PL), and a general-purpose CPU in the system fabricated with the TSMC 7nm technology. 
Designing on Versal ACAP incurs several challenges and we propose AIM: \underline{A}rbitrary-precision \underline{I}nteger \underline{M}ultiplication on Versal ACAP to automate and optimize the design. 
AIM accelerator is composed of AIEs, PL, and CPU. 
AIM framework includes analytical models to guide design space exploration 
and AIM automatic code generation to facilitate the system design and on-board design verification.
We deploy the AIM framework on three different applications, including large integer multiplication (LIM), RSA, and Mandelbrot, on the AMD/Xilinx Versal ACAP VCK190 evaluation board.
Our experimental results show that compared to existing accelerators, AIM achieves up to 12.6x, and 2.1x energy efficiency gains over the Intel Xeon Ice Lake 6346 CPU, and NVidia  A5000 GPU respectively, which brings reconfigurable computing the \emph{most energy-efficient} platform among CPUs and GPUs. 
\end{abstract}


%% file: 1_intro.tex
\vspace{-15pt}
\section{Introduction}
\label{sec:intro}






\begin{figure}
\centering
\includegraphics[width=1\linewidth]{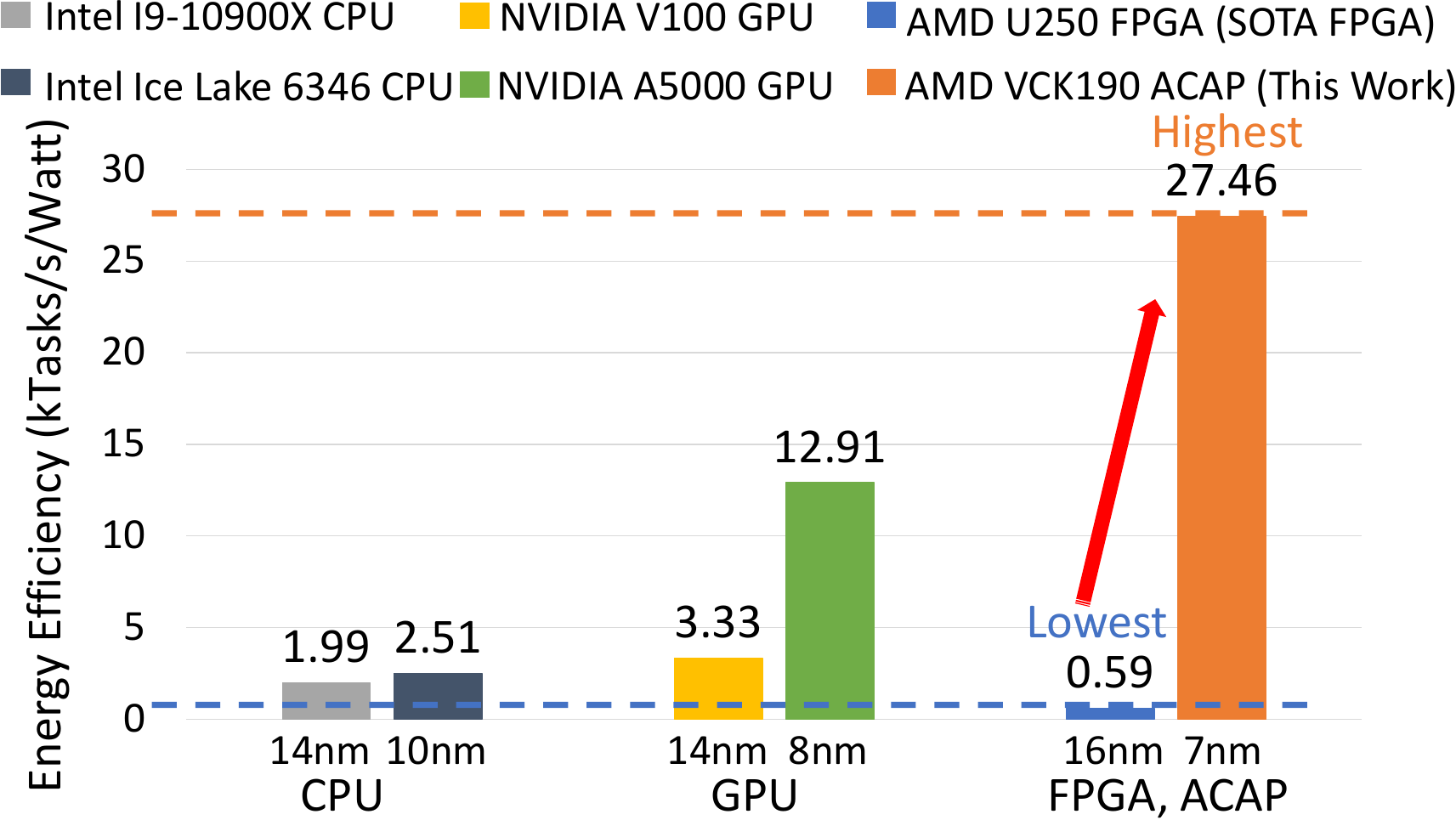}
\caption{Energy efficiency in kTasks/Watt comparison between the prior and the current generation of CPUs, GPUs, and FPGA/ACAP.}
\label{fig:motivation_example}
\end{figure}

Abitrary-precision or large integer multiplication (e.g., $\geq$1024 bit) is one of the most important arithmetic operations for scientific big data analysis~\cite{bailey2012high} (e.g. $\pi$, dispersion coefficients, n-body system) and high-security level data encryption and decryption (e,g. RSA~\cite{rivest1978method,nist}, ECC~\cite{mehrabi2020elliptic}). 
According to ``Application-Specific Key
Management Guidance'' report from National Institute of Standards and Technology (NIST)~\cite{nist}, the highest security level RSA key size is
15,360-bit. As the computational power keeps growing, a larger key
size RSA will be necessary.
Large-bit integer multiplication (LIM) cannot be directly deployed on modern computing platforms which have fixed and lower-bit (32/64-bit) precision function units. 
One common solution is adopting Karatsuba~\cite{weimerskirch2006generalizations}, Schoolbook~\cite{fccm_impress} algorithms, and their variants~\cite{mera2020time} to decompose the high-precision data into smaller segments or limbs, therefore, generating a set of smaller-bit multiplication that can be mapped onto the native hardware intrinsics, e.g., CPU vector instruction units, GPU CUDA cores, FPGA digital signal processing (DSP) blocks.
Among these computing platforms, reconfigurable computing, e.g., FPGA accelerators~\cite{cong2010customizable,cong2022fpga,zhang2018caffeine,chi2022democratizing} are promised to achieve a balance between energy efficiency and flexibility, i.e., better energy efficiency than general-purpose CPUs while preserving flexibility from programmable logic and interconnects. 
We implement the state-of-the-art (SOTA) FPGA LIM accelerator IMpress~\cite{fccm_impress} on the AMD/Xilinx 16nm Ultrascale+ U250 FPGA board~\cite{u250} and compare it with two generations of CPUs (14nm and 10nm) and GPUs (14nm and 8nm) with SOTA CPU/GPU libraries to compute 32768-bit unsigned integer multiplication. 
We calculate the energy efficiency in thousand tasks per second per watt (kTasks/s/Watt) for each computing platform and plot the bar chart.
As shown in Figure~\ref{fig:motivation_example}, in terms of energy efficiency, we observe that IMpress on 16nm FPGA has the \emph{lowest energy efficiency}, i.e., 0.29x of the 14nm CPU and 0.17x of the 14nm GPU. 
Therefore, one key question arises: \emph{Where do the energy efficiency gains of CPUs and GPUs over FPGAs come from?}

We first perform detailed profiling and analyze the energy efficiency gains of the CPUs and GPUs over FPGAs.
By using Intel Software Development Emulator (Intel SDE)~\cite{intelSDE} and Intel Vtune Profiler~\cite{vtune}, we find that over 70\% instructions are vector instructions vmul and vadd on Intel 14nm 19-10900X CPU~\cite{intelI9} (Cascade Lake, Intel 2nd Generation Scalable Processor).
In Cascade Lake CPU, there are two 512-bit Advanced Vector Extensions (AVX-512) instruction units in each core~\cite{intelManual}.  
Each vector instruction performs 8 lanes of computation with each lane computing up to 64-bit operands to accelerate performance. 
Intel 10nm Xeon 6346 CPU~\cite{intel6346} (Ice Lake, Intel 3rd Gen.) has a similar architecture added with AVX-512 Integer Fused Multiply-Add (AVX512\_IFMA)~\cite{papazian2020new} instructions and the profiling results show that about 78\% of instructions being vector AVX512\_IFMA instructions.
GPU 14nm Volta GPU V100~\cite{nvidiaVolta} and 8nm Ampere GPU A5000~\cite{nvidiaAmpere} also feature abundant vector units. 
In Volta and Ampere GPU, there are four CUDA cores in each streaming multiprocessor (SM), and each CUDA core is capable of executing 16 INT32 operations per clock.
In contrast, FPGA uses byte-level computation block DSPs and bit-level lookup tables (LUTs) to compose the needed coarse-grained larger-bit computation module and needs to pay the control overhead for every single module.
With the dedicated vector units, CPUs and GPUs execute the same instruction for multiple data lanes, therefore, spend less energy in control logic, i.e., instructions, and this explains the energy efficiency gains of the CPUs and GPUs over FPGAs in LIM as SOTA CPU/GPU libraries decompose LIM into 32/64-bit multiplications and summations that are efficiently mapped to the dedicated vector units on the CPUs and GPUs.
A follow-up question arises: \emph{Can reconfigurable computing do better if with vector units?}

Our answer is ``Yes". 
In this paper, we propose to map arbitrary-precision integer multiplication onto such a ``FPGA+vector units'' platform, i.e., AMD/Xilinx Versal ACAP architecture~\cite{xilinxAcap}, a heterogeneous reconfigurable computing platform that features 400 AI engine tensor cores (AIE) running at 1 GHz, FPGA programmable logic (PL), and a general-purpose CPU in the system fabricated with the TSMC 7nm technology. 
Designing on Versal ACAP incurs \emph{new challenges}: \textbf{First}, how to decompose the large-bit integer multiplication onto smaller-bit computation modules and map them onto AIEs, PL, and CPU on Versal ACAP? 
\textbf{Second}, how to decide the parallelism within a single accelerator kernel and how to perform resource allocation among multiple accelerators to achieve the optimal system throughput? 
\textbf{Third},  how to integrate the accelerator in end-to-end real-world applications that have different kernels? 
\textbf{Fourth}, can we automate the design process and reduce the programming efforts for the
system implementation?

To solve the challenges and answer the research questions, we propose the AIM architecture and its automation framework, the AIM framework.
Our contributions are summarized below:
\begin{itemize}[leftmargin=*]
\item \textbf{AIM Systematical Design Methodology and AIM Architecture}:
In Section~\ref{sec:aimAccDesign}, we propose a thorough design methodology including workload partition and AIM architecture featured with four-level dataflow to accelerate arbitrary-precision integer multiplication on Versal ACAP. 
To the best of our knowledge, AIM is the first accelerator for this domain on Versal ACAP.

\item \textbf{AIM Design Automation Framework}:
In Section~\ref{sec:aimFramework}, we introduce the AIM framework which includes analytical models to guide design space exploration 
and AIM automatic code generation to facilitate the system design and on-board design verification.
We also show how to deploy the AIM framework and integrate AIM accelerators in three different applications, including large integer multiplication (LIM), RSA, and Mandelbrot, on the AMD/Xilinx Versal ACAP VCK190 evaluation board.

\item 
Our on-board experiments in Section~\ref{sec:experiments} show that compared to SOTA accelerators and libraries, AIM achieves up to 46.7x, 12.6x, and 2.1x, energy efficiency gains over FPGA accelerator IMpress on AMD/Xilinx 16nm Alveo U250, Intel 10nm Ice Lake 6348 CPU, and NVidia 8nm A5000 GPU.

\item \textbf{AIM Open-Source Tools}: 
We open-source our tools with a detailed step-by-step guide to reproduce all of the results presented in this paper and for others to learn and leverage AIM in their end-to-end applications:
\textbf{\emph{\url{https://github.com/arc-research-lab/AIM}}}.


\end{itemize}

%% file: 2_related.tex
\vspace{-5pt}

\section{Related Work}
\vspace{-5pt}
\begin{figure*}
\centering
\includegraphics[width=0.85\linewidth]{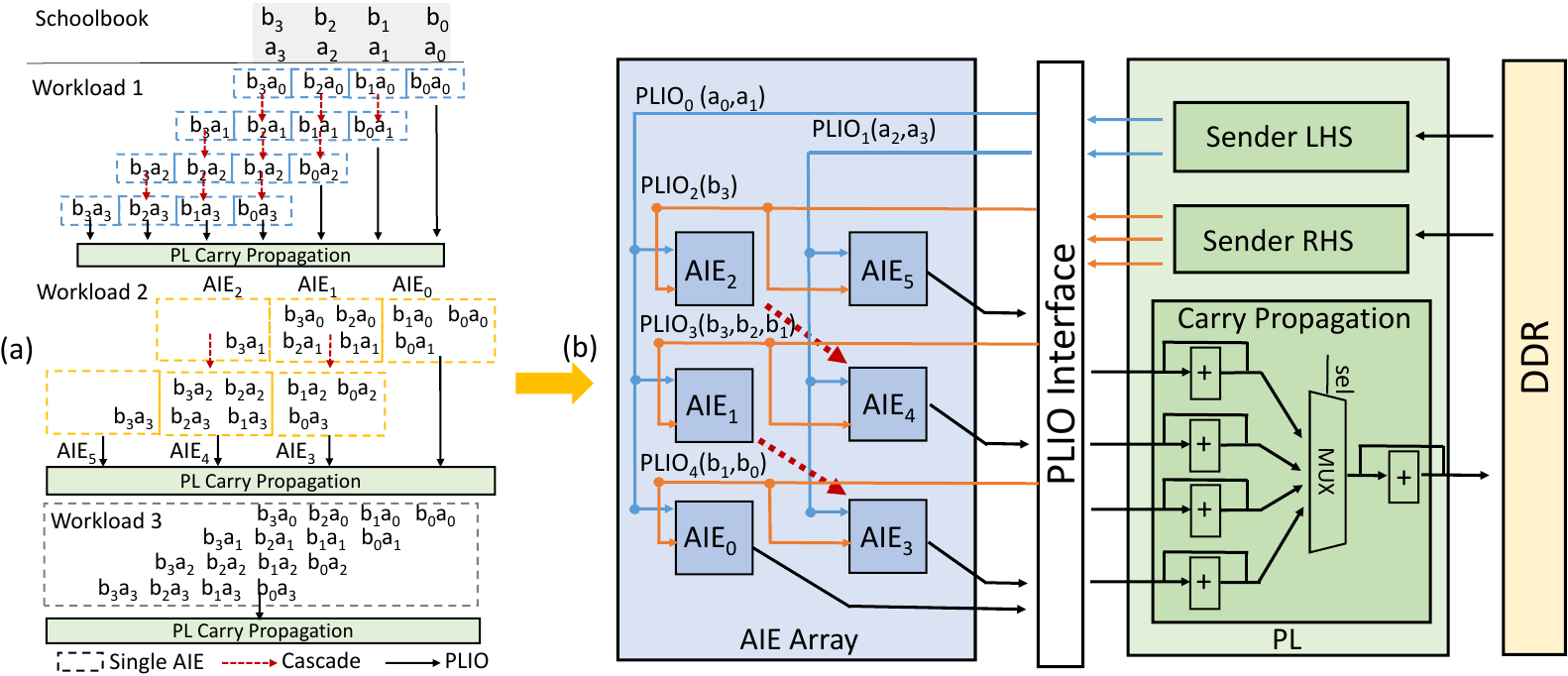}
\vspace{-5pt}
\caption{{(a) Different workload partition schemes based on Schoolbook algorithm; (b) AIM architecture overview.}}
\label{fig:aim_arc}
\vspace{-20pt}
\end{figure*}

In this section, we discuss different decomposition methods and existing accelerators and libraries for large integer multiplication on various platforms, including CPU, GPU, FPGA, and ASIC. 
\vspace{-5pt}
\subsection{Decomposition Methods}
By adopting decomposition methods, the two operands of a large multiplication are decomposed into smaller limbs and can be calculated using smaller multipliers in parallel. The Schoolbook decomposition is as follows:

\vspace{-15pt}
\begin{eqnarray}
\small{
\begin{gathered}
\label{eq:schoolbook}
opA = opA_{h}:opA_{l} \\
opB = opB_{h}:opB_{l} \\
opA * opB = \underbrace{(opA_{h} * opB_{h})}_{\blacksquare} << n + \textcolor{cyan}{\underbrace{\textcolor{black}{(opA_{h} * opB_{l})}}_{\bullet}} << \frac{n}{2} \\ 
+\textcolor{purple}{\underbrace{ \textcolor{black}{(opA_{l} * opB_{h})}}_{\bigstar}} << \frac{n}{2} + \textcolor{brown}{\underbrace{\textcolor{black}{(opA_{l} * opB_{l})}}_{\pentagram}}
\end{gathered}
}
\end{eqnarray}

The key idea of Schoolbook decomposition is to decompose the operands into two parts (e.g., a 64-bit $opA$ into higher 32-bit as $opA_{h}$ and lower 32-bit as $opA_{l}$) and perform four partial products: $opA_{h}*opB_{h}$, $opA_{l}*opB_{l}$, $opA_{h}*opB_{l}$, $opA_{l}*opB_{h}$ followed by summations.
Although the computation complexity of Schoolbook decomposition is $O(N^2)$ with N as the number of bits, and the largest one among existing decomposition algorithms, Schoolbook decomposition is hardware-friendly and has been selected to build the fundamental compute block (base-case) in existing libraries such as GNU Multiple Precision
Arithmetic Library (GMP)~\cite{cpu_gmp} and MPApca~\cite{cambricon_p}. 
Karatsuba~\cite{weimerskirch2006generalizations} and Toom-Cook~\cite{mera2020time,cook1969minimum} decomposition algorithms introduce more additions to the partial results of smaller limbs to decrease the total number of multiplication needed. 
Equation~\ref{eq:Karatsuba} shows that Karatsuba performs three partial products: 
$opA_{h}*opB_{h}$, $opA_{l}*opB_{l}$, $(opA_{h}+opA_{l})*(opB_{h}+opB_{l})$. However, Karatsuba needs more temporary storage to reuse the already-computed partial products and introduces three extra summations. 
Toom-Cook decomposition splits operands into more limbs and applies more complicated arrangements. 

\vspace{-12pt}
\begin{eqnarray}
\small{
\begin{gathered}
\label{eq:Karatsuba}
opA * opB = \underbrace{(opA_{h} * opB_{h})}_{\blacksquare} << n + \textcolor{cyan}{\underbrace{\textcolor{black}{(opA_{l} * opB_{l})}}_{\bullet}} + \\
 (\textcolor{purple}{\underbrace{\textcolor{black}{(opA_{h} + opA_{l}) * (opB_{h} + opB_{l})}}_{\bigstar}} - \\
 \underbrace{(opA_{h} * opB_{h})}_{\blacksquare} - \textcolor{cyan}{\underbrace{\textcolor{black}{(opA_{l} * opB_{l})}}_{\bullet}} ) << \frac{n}{2}\\
\end{gathered}
}
\end{eqnarray}

Therefore, those decomposition algorithms have smaller complexity. 
However, this trick entails a larger memory footprint to store the partial results than Schoolbook decomposition. As reported in \cite{cambricon_p}, decomposing one 1,000,000-bit multiplication into 32-bit multiplications requires 1.72 GB of storage and the memory footprint can be smaller if a larger multiplier is available (1024-bit multiplier requires 223.71 MB storage in this case). 

\vspace{-5pt}
\subsection{Prior Accelerators and Libraries}


\noindent\textbf{CPU.}
The GMP~\cite{cpu_gmp} is one of the most popular high-performance libraries for CPUs to compute arbitrary precision arithmetic. 
Some work ~\cite{cpu_avx, cpu_avx2, cpu_avx3} utilize Intel's Advanced Vector Extensions to efficiently compute large integer multiplication on the CPU. 
GMP adopts Schoolbook decomposition as its base-case multiplication (up to 2048-bit) and selects other decomposition methods for large-bit multiplications on base-case multipliers. 

\noindent\textbf{GPU.}
GPUs also rely on software libraries to compute large multiplications. Cooperative Groups Big Numbers (CGBN)~\cite{gpu_cgbn} is a general solution for GPU that utilizes CUDA cores to realize high parallelism. However, CGBN only supports up to 32k bits multiplication, and for smaller sizes, the operands must be evenly divisible by 32. Dieguez et al.~\cite{gpu_adrian} adopts the Strassen FFT algorithm and a divide-and-conquer algorithm to efficiently compute large integer multiplication on GPU. Goey et al. ~\cite{gpu_ntt_he} accelerate large integer multiplication on GPU using NTT and  apply it to a homomorphic encryption scheme.

\noindent\textbf{FPGA.}
On FPGA, users can directly use vendor tools, for example, on AMD/Xilinx FPGAs, users can compute multiplications up to 2048-bit directly using HLS~\cite{vitis_security}.
Langhammer et al.~\cite{folded_intmul} proposes an efficiently folded multiplier using Karatsuba decomposition. IMpress~\cite{fccm_impress} designs HLS-based~\cite{vitis_hls} FPGA accelerators, combines Karastuba and Schoolbook decomposition methods, and adopts equality saturation to balance the hardware resource utilization. 
Vitali et al.~\cite{fpga_date} combine Karastuba and Comba decomposition to generate a throughput-oriented multiplier design on FPGA. 

\noindent\textbf{ASIC.}
Cambricon-P~\cite{cambricon_p} is an efficient ASIC for arbitrary integer computing, and its base-case hardware multiplier (up to 32768-bit) is based on Schoolbook decomposition and aligns the partial results to enable fast carry propagation. 
Similar to GMP, for larger multiplication, Cambricon-P is able to choose different decomposition methods on base-case multipliers. 
Mert et al.~\cite{asic_mm} design a low-latency large integer modular squaring ASIC.

\noindent\textbf{ACAP.} Prior works have proposed accelerators on ACAP for deep learning ~\cite{fpga23acap, dac23acap}, graph neural network~\cite{fpl22acap_gcn}, stencil computation~\cite{ics23acap_stencil, fpga23acap_stencil}, etc. To the best of our knowledge, we are the first work to implement arbitrary-precision integer multiplication on ACAP. We use Schoolbook decomposition for hardware-friendly mapping and we leave the other decomposition methods as future work.

%% file: 3_versal_architecture.tex
\vspace{-2pt}
\section{Versal ACAP Architecture Overview}
\vspace{-3pt}
\label{sec:arch_overview}
In this section, we introduce the overall architecture of the Versal heterogeneous SoC platform and the AIE Array of the Versal ACAP.
\vspace{-15pt}
\subsection{Versal ACAP Architecture}
Versal ACAP is a computation platform with high performance and high heterogeneity. 
As shown in Figure~\ref{fig:versal_arc}, it is composed of  scalar engines (CPU) for general-purpose processing, programmable logic providing bit-level flexibility, and AI Engines (AIEs) optimized for computation-intensive processing. 
Versal ACAP adopts the multi-level scratchpad memory hierarchy in PL and AIE including the 20 MB SRAM in PL and 12.8 MB local memory in AIE. 
The data in PL SRAM storage can be shared with all the AIEs through the interface connections between PL and AIE, namely PLIO.

\vspace{-5pt}
\subsection{AIE Array}
We highlight the data movement and computation of the intelligent engines (AIEs) in Figure~\ref{fig:versal_arc}. 
Each AIE is a very long instruction word (VLIW) supported vector processor that runs at 1 GHz. 
In each cycle, it supports up to 7-instruction parallelism including 2 loads, 1 store, 1 vector operation, 1 scalar operation, and 2 move operation. For our target device VCK190, there are 400 AIEs and physically form an 8 rows $\times$ 50 columns array. The AIE array applies a tiled architecture in that each AIE owns a 32KB local memory, 2Kb vector register, and 3Kb accumulation register. The local memory of AIE can be shared with the neighboring 4 AIEs through the 256bits/cycle high bandwidth connections and with the non-neighboring AIEs through the 32bits/cycle AXIS stream connections. Between the neighboring AIEs in the same row, a dedicated cascade connection enables fine-grained data transmission from the accumulation registers.
\vspace{-5pt}

%% file: 4_AIM_single_kernel.tex
\begin{figure}
\centering
\includegraphics[width=0.8\linewidth]{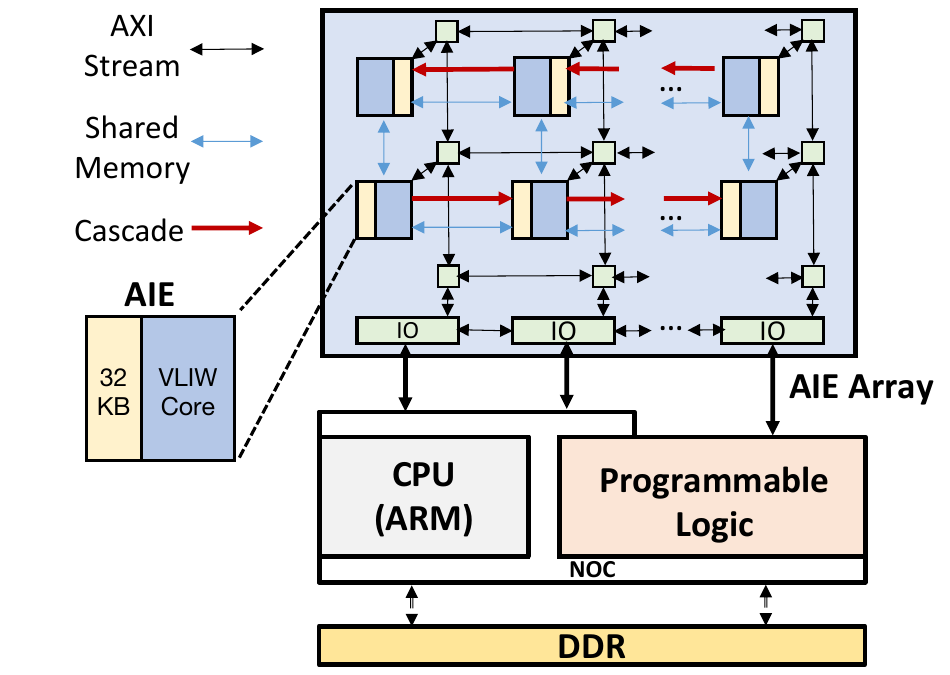}
\vspace{-5pt}
\caption{Versal architecture overview.}
\vspace{-15pt}
\label{fig:versal_arc}
\vspace{-5pt}
\end{figure}
\section{AIM Single Accelerator Design}
\label{sec:aimAccDesign}
In this section, we first introduce the schoolbook decomposition algorithm and analyze the challenges of designing the high throughput accelerator when mapping this algorithm on ACAP. 
We then provide the AIM dataflow architecture overview and our mapping strategy on Versal ACAP.
We also elaborate on the AIM architecture details from the single AIE optimization, scaling out to the AIE array, to the PL fully pipelined carry propagation module.

\subsection{Workload Partition Based on Schoolbook Algorithm}

When dealing with arbitrary size integer multiplication, the schoolbook decomposition algorithm serves as the building block for its relatively low storage demand compared with other decomposition methods, e.g., Karatsuba.
In the decomposition, the large integers can be evenly separated into multiple smaller segments  at a certain granularity. 
Taking a 128-bit multiplication as an example, in Figure~\ref{fig:aim_arc}(a), the two operands are divided into four 32-bit segments. 
Sixteen multiplications are needed to generate the partial results which have $\mathcal{O}(N^2)$ complexity. The final result is obtained by accumulating the partial results within the same column and propagating carry bits to the next column. 
While the large integer multiplication provides good parallelism, a large amount of \textbf{temporal memory footprint} and the \textbf{long carry chain computation} make it non-trivial to design.


\subsection{AIM Overall Architecture}


 Figure~\ref{fig:aim_arc}(b) shows the overview of our proposed AIM architecture, which is composed of the AIE array, PL data processing modules, and the corresponding I/Os. 
 The sub-multiplications in Figure~\ref{fig:aim_arc}(a) can be grouped with different workload partition schemes and mapped onto different numbers of AIEs. 
 Here we use workload 2 to illustrate. 
 The 16 multiplication are partitioned into 6 groups and mapped to 6 AIEs (AIE0-AIE5).  
 The AIEs with read-after-write (RAW) dependency (AIE1$\rightarrow$AIE3, AIE2$\rightarrow$AIE4) are connected with the cascade stream that passes the temporal results in a fine-grained manner. To explore the PLIO reuse, the input data on the same row or in the same hypotenuse direction will be broadcast by the senders on the PL side via the PLIO interface.
 In order to overlap the long latency caused by the carry chain, AIM takes advantage of the flexibility of programmable logic on ACAP and designs a dedicated high-throughput fine-grained carry propagation module.


\subsection{AIM Four-Level Dataflow of AIM Architecture}
\setlength{\textfloatsep}{0pt}

\begin{figure}
\begin{lstlisting}[language=C, label=alg:aim_pl_dataflow,caption=Data tiling and dataflow in AIM.]
L3: PL_load_input_data_from_DDR(...);
L2: data_preprocessing_on_PL(...);
L1: // Parallel computation in AIE array
    for(int c = 0; c < AIE_COL; c++):
    // Dependency exists on different rows
    for(int r = 0; r < AIE_ROW; ++r):
L0:    // Single AIE compute flow
       for(int w = 0; w < B_W/P_W; ++w):
       for(int h = 0; h < A_H/P_H; ++h):
           vector_mul(...); //call packed instr.
L2: carry_propagation_on_PL(...);
L3: PL_store_results_DDR(...);
\end{lstlisting}
\end{figure}


Listing~\ref{alg:aim_pl_dataflow} shows the pseudo-code of the four-level dataflow:

\noindent\textbf{L0: Single AIE Level (Lines 7-9).} 
In a single AIE level, each AIE/tile computes with carefully designed and packed vector intrinsics instructions.

 \noindent\textbf{L1: AIE Array Level (Lines 4-6).} 
 The grouped tiles are distributed to multiple AIEs computed in parallel. 
 Parallel loop Line 6 shows that for AIEs within the same column, the partial results need to be accumulated and these AIEs are connected using the cascade stream. Parallel loop Line 4 describes AIEs in different columns. The AIE array size $AIE\_COL \times AIE\_ROW$ is determined by the input size and tile granularity in Figure~\ref{fig:aim_arc}(a). 
 For Workload 2, the AIE array size is 3 $\times$ 2. 

\noindent\textbf{L2: PL Data Processing Level (Lines 2\&11).} 
On the PL side, we design multiple stream-based data processing modules. 
By applying a fine-grained sending and receiving strategy, the dedicated data pre-processing and carry propagation modules can keep pace with the throughput of the AIE array and hide the latency of carry propagation.

 
 \noindent\textbf{L3: Off-Chip Level (Lines 1\&12).} 
At the last level, data will be streamed between the DDR and the BRAM on the PL side.

\subsection{Single AIE Kernel Optimization}
The simplified single AIE level computation flow is shown in Listing~\ref{alg:aie_krnl}. The kernel takes two local memory pointers ($in0$, $in1$) and one cascade stream ($acc\_in$) as input (Line 1). 
The input data will be loaded from the local memory or cascade stream into the local vector registers as shown in Lines 4-6. 
Then multiple SIMD instructions are packed together in the $vector\_mul$ function to process the vector registers (Lines 11-20). 
In order to explore the instruction-level parallelism, AIM inserts the load instructions (Lines 15 \& 16) with multiplication instructions to hide the latency for preparing the data needed in the next iteration. 
The output stationary dataflow is used to avoid frequent vector eviction. 
The results will only be sent to the output stream and passed to the next tile/AIE after finishing all the reductions in this tile (Line 9). 
On the AIM architecture, the accumulator is up to 80-bit and the result segments  in AIM are $31bits \times 31bits = 62 bits$. Therefore, the accumulator register is safe to sum up $2^{17}$ partial results with 1 sign bit left.

\begin{figure}
\begin{lstlisting}[language=C, label=alg:aie_krnl,caption=Optimized AIM kernel compute flow.]
AIE_Krnl(in0, in1, out, acc_in):
    for(int w = 0; w < B_W / P_W; ++w):
        // Read partial results from previous AIE
        v8acc80 = read_acc(acc_in) 
        v8a = read(in0) // Read new segment A
        v16b = read(in1)  // Read new segment B
        for(int h = 0; h < A_H / P_H; ++h):
            vector_mul(v8a,v16b,v8acc80)
        write_acc(out, v8acc80)
// carefully pack instructions here:
vector_mul(v8a,v16b,v8acc80):
   v8acc80 += v16b[0:7] * v8a[0]
   v8acc80 += v16b[1:8] * v8a[1]
   v8acc80 += v16b[2:9] * v8a[2]
   v16b_next = read(in1) // load instruction
   v8a_next = read(in0) // load instruction
   ...
   v8acc80 += v16b[7:15] * v8a[7]
   v16b = v16b_next
   v8a v8a_next
\end{lstlisting}
\end{figure}

\subsection{Scaling Out to AIE Arrays}
To achieve the highest system-level throughput, more AIEs should be utilized. In AIM, we adopt a spatial computing fashion.
In this spatial computing, we also exploit the data broadcasting mechanism to reduce the PLIO demand. 
Still, take the AIE array of workload 2 in Figure~\ref{fig:aim_arc}(a) as an example, the AIEs within the same row share the same segments  from operand A, and AIEs aligned in the hypotenuse direction share the same segments  from operand B.
In this case, only 5 input PLIOs and 4 output PLIOs are needed instead of using 12 input PLIOs and 6 output PLIOs. 
The PLIO saving is more significant when mapping to a larger AIE array. 

The cascade stream connects AIEs with the RAW dependency to make better use of the accumulator register and reduce the amount of data that needs to be streamed out to PL for reduction. 
Although this introduces dependencies in the AIE array, the performance is not hurt as we adopt the fine-grained pipeline to minimize the transmission overhead. 
In Listing~\ref{alg:aie_krnl}, each AIE first calculates multiplications that need to be accumulated together. 
Then, it transmits the partial results to the next AIE at line 9 and starts accumulating on another register for the rest multiplications. 
In this way, both AIEs can start computing earlier, and their computation timelines largely overlap. 

\begin{figure}
\centering
\includegraphics[width=1\linewidth]{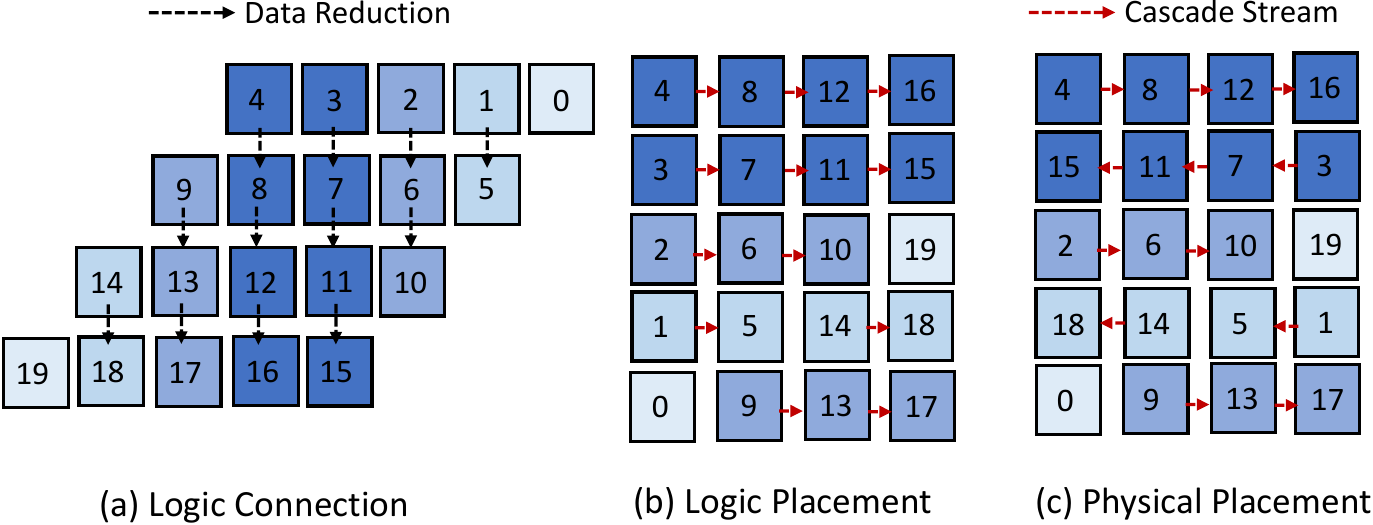}
\vspace{-15pt}
\caption{Three AIE placement stages to map an AIM accelerator with 20 AIEs: (a) AIEs in the logic computation; (b) placement considering the 2D rectangle shape of the AIE array; (c) physical placement considering the cascade stream directions in the AIE array.}
\label{fig:placement_algo}
\vspace{-13pt}
\end{figure}

\begin{figure}
\centering
\includegraphics[width=0.6\linewidth]{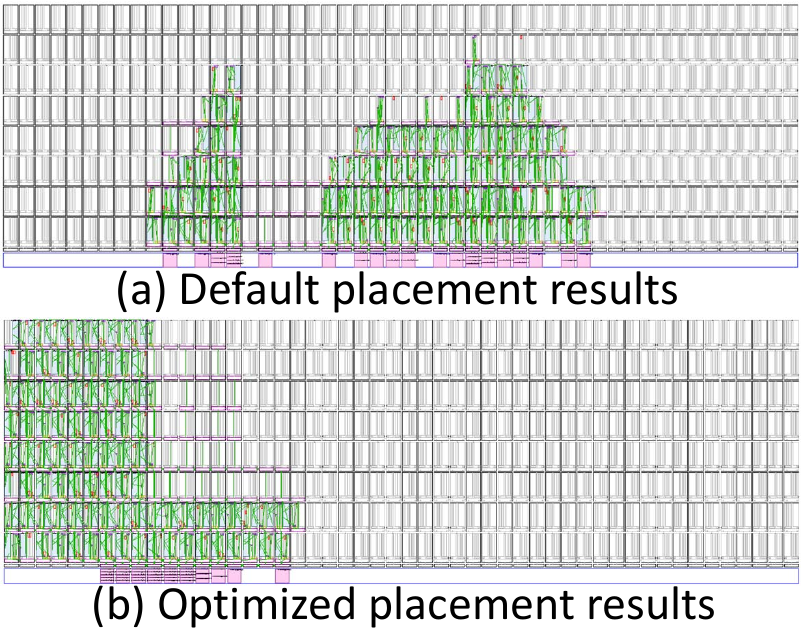}
\vspace{-5pt}
\caption{AIE placement before vs. after two-step placement algorithm.}
\label{fig:aie_placement}
\vspace{-5pt}
\end{figure}

\noindent\textbf{AIE Placement Optimization.}
When scaling out to multiple AIEs, both logical connection and physical connection constraints should be fulfilled. For logical connection, shown in Figure~\ref{fig:aim_arc}(b), segments  from operand A are broadcast to AIEs in the same row, segments  from operand B are broadcast along the same column, and the cascade stream connects AIEs in the reduction dimension. This is difficult for the physical connection. 
As mentioned in Section~\ref{sec:arch_overview}, the 400 AIEs on ACAP are distributed in 8 rows and 50 columns, and the cascade stream connects AIEs in the same row. 
Without proper guidance, the search space is huge.  
As shown in Figure~\ref{fig:aie_placement}(a), the vendor-provided default AIE placer tends to occupy the middle part of the AIE array, and the occupied AIEs form an irregular shape. This default placement fails if more AIEs are used. 
Thus, a better placement algorithm is needed. 

We propose a two-step placement algorithm to solve this and enable more AIE usage. 
Figure~\ref{fig:placement_algo}(a) shows the logic connection when mapping the whole computation onto 20 AIEs.
In the first step, we rearrange AIEs in different columns and match them to form AIE links with the same length.
This gives a regular rectangular shape, as shown in Figure~\ref{fig:placement_algo}(b). 
We transpose the AIE link direction from the vertical direction in Figure~\ref{fig:placement_algo}(a) to the horizontal direction in Figure~\ref{fig:placement_algo}(b) to match the physical cascade stream direction as horizontal. 
After the rearrangement, the AIEs in the same link will be placed together. 
In the second step, we find the exact AIEs to put each AIE link. 
We place AIE links and map them onto the physical AIEs starting from the left part of AIE arrays to the right and from the bottom part of the AIE arrays to the up.
Considering the actual cascade streams flow in opposite directions for every adjacent two rows,  Figure~\ref{fig:placement_algo}(c) shows the actual placement to reflect this.
By adopting the proposed placement algorithm, AIM is able to use up to 396 AIEs. 
Figure~\ref{fig:aie_placement}(b) shows the optimized placement results. 
\vspace{-5pt}
\subsection{Specialized PL Data Processing Module Design}
To keep pace with the high throughput of the AIE array and hide the latency caused by the long carry chain, we design dedicated senders and carry propagation processing modules on the PL side.


\noindent\textbf{PL Sender.}
The senders in Figure~\ref{fig:aim_arc}(b) are responsible for doing the bitwidth conversion between DDR and PL as well as exploring the broadcast opportunity for feeding the data to the AIE array. 
To achieve higher off-chip DDR bandwidth, 512 bits data granularity is applied to load the two input operands. 
In contrast, the maximum bitwidth of PLIO (PL $\leftrightarrow$ AIE) is 128-bit. 
A bitwidth conversion module is created inside each sender module. 
Besides, AIEs only support 32-bit data selection granularity with signed integers and lack unsigned 32-bit vector instructions. 
Therefore, the senders need to prepare data for AIE by slicing the large chunks into 31-bit segments and inserting zero as the sign bit. 



\noindent\textbf{PL Carry Propagation.}
The carry propagation module propagates carry bits for the partial results from the AIE array.
In AIM, the multiplication results are first reduced in the accumulator register. 
The carry bits in the accumulator are further reduced on the PL side.
The carry chain can be very long and can easily be the bottleneck of the whole system. 
In AIM, this is avoided by breaking the long carry chain into shorter carry chains and processing the carry in two steps. 
As shown in Figure~\ref{fig:aim_arc}(b), each PLIO is connected to a module to calculate carry bits in 128-bit granularity. 
These carry adders are computing simultaneously.
Then, more bits (512-bit granularity) can be added in one cycle  by using LUTs and flip-flops (FFs) to generate the correct carry bits.

%% file: 5_AIM_framework_application.tex
\vspace{-5pt}
\section{AIM Design Automation Framework and Integration in End-to-end Applications}

In this section, we will discuss the AIM design automation framework. Then, we will use three representative applications to illustrate how to integrate AIM into end-to-end applications.

\label{sec:aimFramework}
\vspace{-5pt}
\subsection{AIM Design Automation Framework Overview}
\begin{figure}
\centering
\includegraphics[width=1\linewidth]{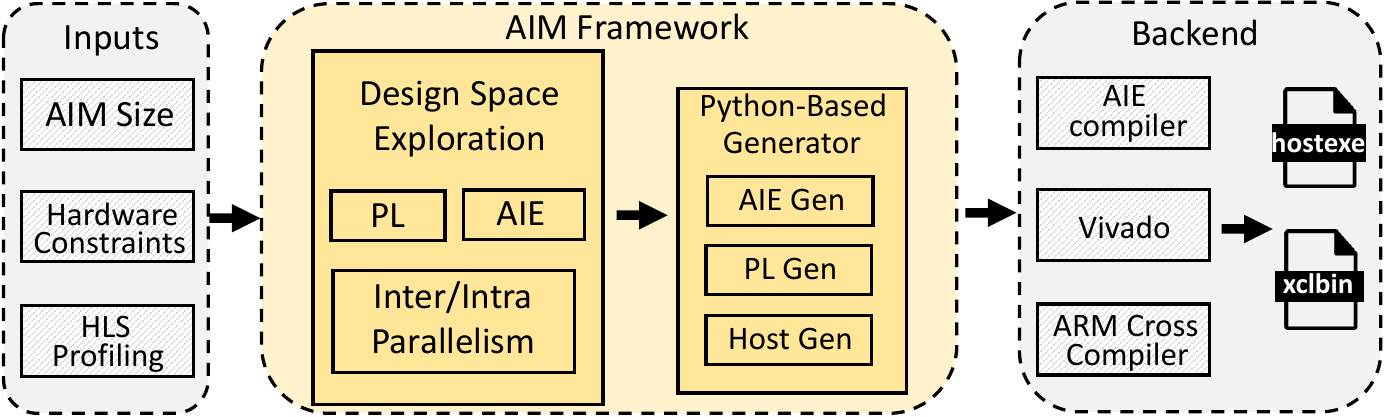}
\vspace{-15pt}
\caption{AIM framework.}
\label{fig:aim_framework}
\end{figure}
Figure~\ref{fig:aim_framework} shows the AIM framework overview. 
The AIM framework generates customized AIM architecture with user-specified input data sizes and hardware resource constraints. 
A one-time HLS profiling results with LUT/BRAM utilization and PL kernel execution time are needed to guide the design space exploration (DSE) in the next step. 
The AIM-DSE module is responsible for searching optimal PL and AIE configurations to create a customized AIM architecture exploiting inter-task and intra-task level parallelism. 
Then, the Python-based AIM automatic code generator (AIM-ACG) takes the optimal configurations as input and emits the code for AIE kernels, AIE array mapping, high throughput PL modules, and CPU host. 
Finally, AIM automatically launches the vendor-provided back-end tools, Vivado~\cite{vitis}, Vitis HLS~\cite{vitis_hls}, AIE Compiler~\cite{aieCompiler}, etc., to generate the hardware configuration bitstream (.xclbin) for PL \& AIE and host executable binary (.hostexe) for ARM CPU on ACAP.

{\color{black}
\vspace{-5pt}
\subsection{AIM Architecture Exploiting Inter-/Intra-task Parallelism}
We consider two different levels of parallelism, inter-task parallelism, and intra-task parallelism as shown in Figure~\ref{fig:parallelism}. 
In this example, every PE calculates an independent task and occupies 9 AIEs. Therefore, the intra-task is 9. 
The AIM architecture is able to accommodate multiple PEs for different tasks. 
In Figure~\ref{fig:parallelism}, N PEs are placed and the inter-task parallelism is N.


The generated sub-multiplications can be calculated in parallel.
As shown in Figure~\ref{fig:aim_arc}(a), the necessary multiplications to be computed form a parallelogram shape, and different workload partition schemes on each AIE lead to different mapping efficiency and AIE kernel efficiency. 
Workload 1 in Figure~\ref{fig:aim_arc}(a) maximizes the intra-task parallelism, and maps the whole computation with 16 AIEs within each task or PE.
However, if the input size is smaller, fewer instructions can be packed which leads to a decreased AIE kernel execution efficiency. 
Workload 3 in Figure.~\ref{fig:aim_arc}(a) maps the whole computation onto a single AIE within each task or PE and maximizes the inter-task parallelism by calling multiple PEs. 
As each task or PE needs corresponding PL resources for the sender modules and the carry propagation modules, more PEs mean more PL resource usage.
We explore the proper intra-task and inter-task parallelism by model-guided DSE and find the ``sweet point'' in the whole DSE space.
Table~\ref{tbl:granularity_compare} shows the system level throughput of the three different parallelism configurations. Case 2 (workload 2) in Figure~\ref{fig:aim_arc}(a) is the optimal design and the total number of AIEs in the system is 210.
Workload 1 occupies more AIEs with fewer PL resources and it is easier to use more total numbers of AIEs (306). 
However, each AIE's efficiency is low in this case.
In workload 3, each AIE tends to consume more programmable logic. Therefore the total number of AIEs can be used (80) is bounded by the PL resource. 
In summary, the AIM architecture is flexible with different design configurations and we will use DSE to guide the search to achieve the optimal system-level throughput.
\vspace{-0.07in}
\begin{figure}
\centering
\includegraphics[width=0.70\linewidth]{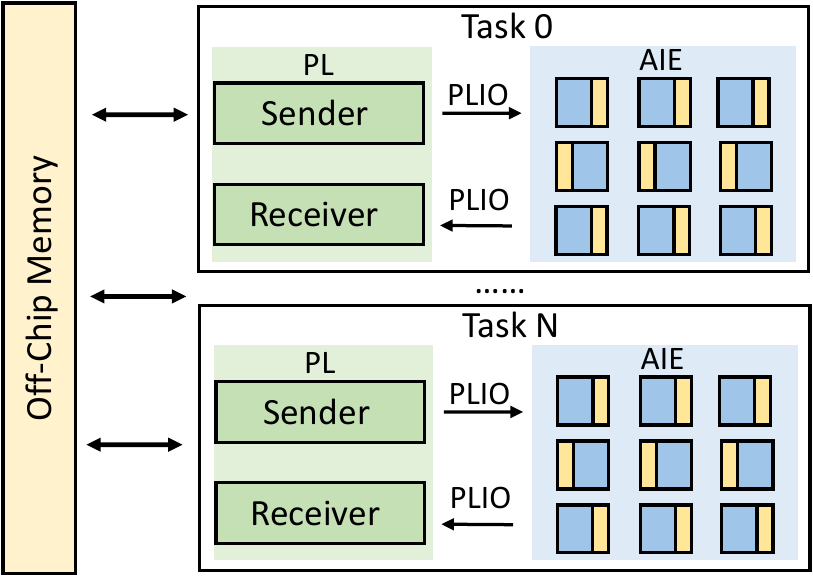}
\vspace{-5pt}
\caption{Inter-task parallelism and intra-task parallelism.}
\label{fig:parallelism}
\vspace{-10pt}
\end{figure}

\input{table/granularity}

\subsection{AIM Analytical Models and Design Space Exploration}
\vspace{-2pt}
\noindent\textbf{DSE Configurable Variables: ($P_{Inter}$, $P_{Intra}$).}
To maximize the overall system throughput, we build the AIM-DSE that takes user-specified data size $N$ and hardware resource constraints $C_{i}$, $i \in \{LUT, RAM, DSP, \#AIE\}$ as inputs. The output of our AIM-DSE is the inter-level parallelism $P_{Inter}$ and intra-level parallelism $P_{Intra}$.



The optimization goal and constraints are summarized as follows:
\vspace{-15pt}
\begin{eqnarray}
\small{
\begin{gathered}
\label{eq:optimization}
\max \;\;\; Throughput(P_{Inter}, P_{Intra}) \\
s.t. \;\; Resource_i(P_{Inter}, P_{Intra}) \leq \mathit{C_i} \\
i = LUT, BRAM, PLIO, AIE
\end{gathered}
}
\end{eqnarray}
\vspace{-10pt}

\noindent\textbf{Overall Throughput Modeling.} Since the AIE array is a 2D array, the $P_{Intra}$ has two dimensions which represent the number of AIEs in AIE array's row and column as shown in Equation~\ref{eq:Para_intra}.
\vspace{-8pt}
\begin{eqnarray}
\small{
\begin{gathered}
\label{eq:Para_intra}
P_{Intra} = P_{Intra0} \cdot P_{Intra1} 
\end{gathered}
}
\end{eqnarray}
\vspace{-20pt}

The number of segments ($S_{0,1}$) for two operands assigned to a single AIE is determined by the input size $N$ and intra-task parallelism $P_{Intra}$: 
\vspace{-12pt}
\begin{eqnarray}
\label{eq:workload}
\small{
\begin{gathered}
S_{0,1} = \frac{ \lceil\frac{N}{31} \rceil }{P_{Intra0,1}}
\end{gathered}
}
\end{eqnarray}
\vspace{-12pt}

where SIMD is the adopted vector parallelism in the single AIE kernel and the 31-bits is set as the segment granularity in Equation~\ref{eq:workload}.

Once the workload is determined, the execution efficiency $Eff$ of the single AIE kernel can be obtained from the cycle-accurate AIE simulator. The AIE compute clock cycle is formulated as follows:
\vspace{-5pt}
\begin{eqnarray}
\small{
\begin{gathered}
\label{eq:aie_cycle}
AIE_{cyc} = \frac{S_0 \cdot S_1}{SIMD \cdot Eff}
\end{gathered}
}
\end{eqnarray}
We characterize the execution time of the sender and carry propagation modules based on the Vitis HLS~\cite{vitis_hls} report. The system's overall throughput can be calculated as follows:
\vspace{-5pt}
\begin{eqnarray}
\small{
\begin{gathered}
\label{eq:obj}
Throughput= \frac {P_{Inter}}{max(Sender_{cyc}, Carry_{cyc}, AIE_{cyc})}
\end{gathered}
}
\end{eqnarray}
\vspace{-10pt}

\noindent\textbf{Hardware Resource Constraints.} The AIE and PLIO consumption need to meet the hardware constraints:
\vspace{-5pt}
\begin{eqnarray}
\small{
\begin{gathered}
\label{eq:pl}
P_{Inter} \cdot P_{Intra} < C_{AIE} \\
(P_{Intra0} \cdot 2 + P_{Intra1} \cdot 2-1) \cdot P_{Inter} < C_{PLIO}
\end{gathered}
}
\end{eqnarray}
\vspace{-10pt}

The consumption of LUT and BRAM are profiled using the Vitis HLS tool and should meet the constraints:
\vspace{-5pt}
\begin{eqnarray}
\small{
\begin{gathered}
\label{eq:pl_res}
LUT_{profile} \cdot P_{Inter} < C_{LUT} \\
BRAM_{profile} \cdot P_{Inter} < C_{BRAM}
\end{gathered}
}
\end{eqnarray}
\vspace{-10pt}
\subsection{AIM Integration into More Complex End-to-end Applications}
Here we use two more complex real-world applications, RSA and Mandelbrot, to demonstrate the integration of AIM into an end-to-end design. 
The advantage of using AIM is the non-multiplication parts can be designed in a pipeline fashion on the PL side, and executed simultaneously with the multiplier.
Therefore, the control flow and other operations' execution time can be hidden with batch processing. 
This explains the reason why AIM achieves higher energy efficiency gains when integrated into end-to-end applications when compared to instruction-based CPUs and GPUs. 

\noindent\textbf{RSA.}
RSA is a commonly used asymmetric cryptographic algorithm that uses different encryption and decryption keys. The security of RSA is based on the mathematical problem of big integer factorizing. The highest security level RSA size in the NIST standard\cite{nist} is 15,360-bit. As the computational power keeps growing, a larger key size RSA  will be necessary. 
The encryption and decryption processes of RSA are shown in Equation~\ref{eq:rsa}.
\vspace{-5pt}
\begin{eqnarray}
\small{
\begin{gathered}
\label{eq:rsa}
Cyphertext = Plaintext^{e_{pub}} \mod \mathit{M} \\
Plaintext = Cyphertext^{e_{prv}} \mod \mathit{M} 
\end{gathered}
}
\end{eqnarray}
\vspace{-10pt}

$M$ is a factor of two large prime numbers ($p, q$), ${e_{pub}}$ and ${e_{prv}}$ satisfy the following conditions:
\vspace{-5pt}
\begin{eqnarray}
\small{
\begin{gathered}
\label{eq:rsa_keys}
\phi = (p - 1)(q - 1)  \\
1 < e_{prv} < \phi  \\
gcd(e_{prv}, \phi) = 1 \\
1 < e_{pub} < e_{prv} \\
e_{pub} \cdot e_{prv} \mod \phi = 1 \\
\end{gathered}
}
\end{eqnarray}
\vspace{-10pt}

The fundamental part of RSA encryption and decryption is modular exponentiation in Equation~\ref{eq:rsa}. 
For fast execution, we adopt exponentiation by squaring and Montgomery Multiplications (MontMul) to reduce the total multiplications needed and avoid slow modular calculation.
RSA encryption is processed in three steps. 
First, the plaintexts and parameters will be read from DDR, and the Montgomery representation of plaintexts will be calculated. 
Second, the RSA modules and MontMul modules perform fast exponentiation and stream data to the AIM architecture. 
Third, the encrypted data exits Montgomery space. 
Considering the side-channel issue in the exponentiation by squaring, AIM calculates Montgomery multiplication regardless of the key value. 

Figure~\ref{fig:RSA_aim} and Figure~\ref{fig:rsa_pipeline} show the RSA dataflow architecture and pipeline of different modules in RSA. 
To fully utilize the AIE array, independent tasks need to be streamed in the AIM architecture. 
AIM reads new tasks and writes computation results simultaneously and it is better for users to decouple the execution of the sender modules and receiver modules via first-in-first-out (FIFO) streams. 
The RSA's pipeline in Figure~\ref{fig:rsa_pipeline} demonstrates the full utilization of AIEs. The key takeaway is that the AIE kernels (3) are fully pipelined and hide the latency of the other kernels (1,2,4,5) that are implemented on PL. Kernels 2 \& 4 are the sender and receiver of MontMul and kernels 1 \& 5 are the sender and receiver of the exponentiation module.
Here we use two independent tasks (denoted $T_0, T_1$) to illustrate. The two tasks are loaded to kernel 1; each task will be sent to kernel 2 twice in each RSA iteration (denoted $M_0, M_1$); Montgomery multiplication requires three multiplications (denoted $S_0, S_1, S_2$). A fine-grained pipeline is designed for PL modules 1, 2, 4, and 5 in Figure~\ref{fig:RSA_aim}. 
In the beginning, kernel 2 reads one multiplication task from kernel 1 and sends it to the AIM architecture. (Time 0)
The AIEs start computing when the first task is completely loaded (Time 1). 
During the computation, AIE architecture keeps reading another new task (Time 1) and prepares it for the computation of the next time step (Time 2). 
In Time 2, the multiplication result of $T_0I_0M_0S_0$ is ready and kernels 4 \& 2 need to prepare multiplication task $T_0I_0M_0S_1$ while computing $T_1I_0M_0S_0$. 
\begin{figure}
\centering
\includegraphics[width=0.8\linewidth]{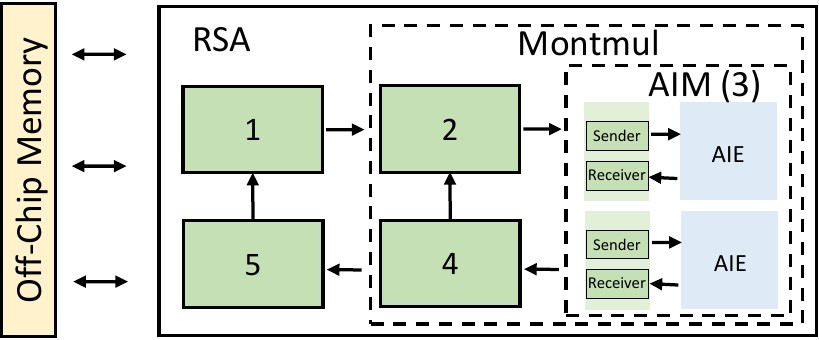}
\vspace{-6pt}
\caption{RSA dataflow architecture.}
\label{fig:RSA_aim}
\vspace{-13pt}
\end{figure}
\begin{figure}
\centering
\includegraphics[width=1\linewidth]{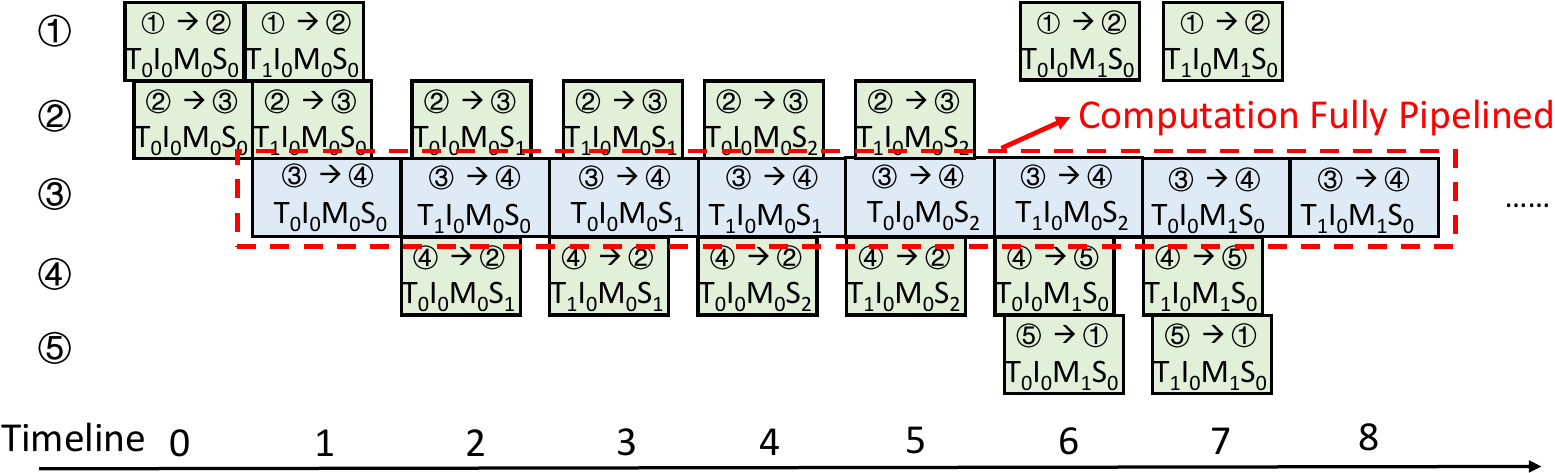}
\vspace{-15pt}
\caption{Pipeline of different modules in RSA. The key takeaway is that the AIE kernels (3) are fully pipelined and hide the latency of the other kernels (1,2,4,5) that are implemented on PL.}
\label{fig:rsa_pipeline}

\end{figure}
\begin{figure}
\centering
\includegraphics[width=.95\linewidth]{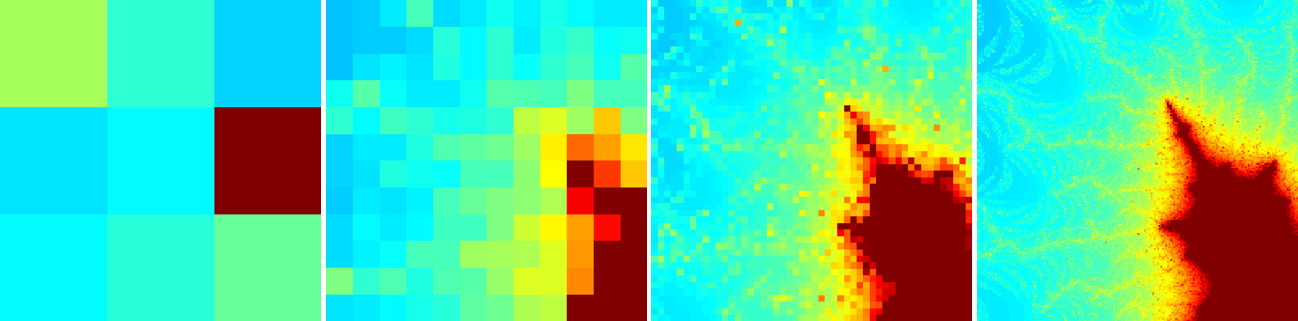}
\vspace{-5pt}
\caption{Plotting the Mandelbrot set from lowest precision (left) to highest precision (right). More precision bits show finer features.}
\vspace{-10pt}
\label{fig:mandelbrot}
\end{figure}

\noindent\textbf{Mandelbrot.}
Mandelbrot set is a type of fractal with detailed structures at arbitrary precision. 
Mandelbrot set is plotted by performing divergence tests, shown in Equation~\ref{eq:mandelbrot}, for sampled points on the complex plane. The divergence tests stop if $|f_c(z)| > 2$ or the iteration number reaches a certain threshold.
The different color shows the number of iterations for this coordinate before stopping.
It has high demands on precision to represent the coordinates since tiny differences in coordinates have significantly different results. 
Figure~\ref{fig:mandelbrot} depicts Mandelbrot set in the same area with the same image size but different precision bits. 
Different from RSA, the number of multiplications cannot be determined ahead of time. The divergence test is performed for every pixel. Different pixels have various iterations before divergence. 
Therefore, this application requires a run-time scheduler. 
AIM takes advantage of programmable logic to implement this run-time scheduler.
\begin{eqnarray}
\small{
\begin{gathered}
\label{eq:mandelbrot}
f_c(0), f_c(f_c(0)), f_c(f_c(f_c(0))), ... \\
f_c(z) = z^2 + c
\end{gathered}
}
\end{eqnarray}

\input{table/Experiment_setup}

\input{table/DSE_Accuracy}

%% file: table/granularity.tex
\begin{table}
\caption{System level performance of 8192-bit multiplier on different parallelism strategies.}
\vspace{-7pt}
\label{tbl:granularity_compare}
\begin{adjustbox}{width=\columnwidth,center}

\begin{tabular}{ c | c | c | c | c | c | c | c}
\toprule
    Case  & $P_{Intra}$ & $P_{Inter}$ & \#bits/AIE & PKT  & LUT &BRAM & Tasks/s \\
\midrule
   1 & 306 & 1   & 496   & 1  & 43.6\% & 4.7\%& 1.6M \\
   2 & 30  & 7   & 1736  & 1  & 78.1\% & 16.7\%& 9.6M \\
   3 &  1  & 80  & 8192  & 4  & 60.5\% & 98.6\%& 5.7M  \\
\bottomrule
\end{tabular}
\end{adjustbox}
\vspace{-3pt}
\end{table}

%% file: table/Experiment_setup.tex
\begin{table}
\begin{center}
\caption{Experiment Setup for CPU and GPU.}
\vspace{-5pt}
\label{tbl:exp_setup}
\begin{adjustbox}{width=.9\columnwidth,center}
    \begin{tabularx}{\linewidth}{ |c| *{2}{>{\centering\arraybackslash}X|}}

    \hline
    \multirow{5}{*}{CPU} & Type & Intel Xeon Gold 6346  \\
    \cline{2-3}
    & Fabrication   & 10nm  \\
    \cline{2-3}
    & Frequency   & 3.1GHz \\
    \cline{2-3}
    & TDP   & 205W/CPU  \\
    \cline{2-3}
    & Library   & GMP Version 6.2.1  \\
    \hline
    \multirow{5}{*}{GPU} & Type & NVIDIA A5000 \\
    \cline{2-3}
    & Fabrication   & 8nm  \\
    \cline{2-3}
    & Frequency   & 1.17GHz  \\
    \cline{2-3}
    & TDP   & 230W  \\
    \cline{2-3}
    & Library   & CGBN Version 2.0 \\
    \hline
    \end{tabularx}
\end{adjustbox}
\end{center}
\vspace{-20pt}
\end{table}

%% file: table/DSE_Accuracy.tex
\begin{table}
\vspace{5pt}
\caption{Model VS. on-board measured performance (Tasks/s) for 65,536-bit LIM on AIM. PL frequency is reported in MHz.}
\vspace{-5pt}
\label{tbl:dse_accuracy}
\begin{adjustbox}{width=\columnwidth,center}
\begin{tabular}{ c | c | c | c | c | c | c }
\toprule
   $P_{Intra}$ & $P_{inter}$ &\#bits/AIE & Freq. & Model & On-board & Error \\
\midrule
    20           & 8 & 16616  & 175 & 185.7k & 186.2k & 0.3\% \\
    30           & 7 & 13144  & 176 & 255.5k & 256.2k & 0.3\% \\
    42           & 6 & 11160  & 184 & 299.6k & 302.2k & 0.8\% \\
    56           & 5 & 9424   & 190 & 344.4k & 348.3k & 1.1\% \\
    72           & 4 & 8432   & 190 & 340.0k & 344.5k & 1.3\% \\
    90           & 4 & 7440   & 186 & 430.1k & 436.6k & 1.5\% \\
    110          & 3 & 6696   & 207 & 392.6k & 399.1k & 1.7\% \\
    \textbf{132} & \textbf{3} & \textbf{6200}   & \textbf{209} & \textbf{452.8k} & \textbf{459.8k} & \textbf{1.5\%} \\
    156          & 2 & 5704   & 206 & 352.1k & 356.5k & 1.3\% \\
    182          & 2 & 5208   & 207 & 415.9k & 387.3k & -6.9\% \\
    210          & 1 & 4712   & 206 & 249.4k & 254.5k & 2.0\% \\
    272          & 1 & 4216   & 208 & 280.3k & 270.6k & -3.5\% \\
\bottomrule
\end{tabular}
\end{adjustbox}
\vspace{-4pt}
\end{table}

%% file: 6_Experiments.tex
\vspace{-13pt}
\section{Experiments Results}
\vspace{-5pt}
\label{sec:experiments}
\input{table/single_kernel_comparison}

In this section, we 
report the performance, and energy efficiency of the AIM designs from on-board measurement, 
demonstrate the accuracy of our analytical model,  and compare AIM designs with other platforms including CPUs, GPUs, FPGA, and ASIC. 
\vspace{-5pt}
\subsection{Experimental Setup}
\vspace{-2pt}
We evaluate AIM designs on AMD/Xilinx Versal VCK190 board~\cite{versal_vck190}, and we use Vitis 2021.1 for system implementation. 
All AIEs are running at 1 GHz and the PL modules' frequency is the maximal achievable frequency after implementation. 
We use AMD/Xilinx board evaluation and management tool~\cite{BEAM}
to measure the power of VCK190 during the execution.
We compare AIM designs with state-of-the-art arbitrary integer multiplication libraries on CPU and GPU, and the CPU and GPU setup is summarized in Table~\ref{tbl:exp_setup}. 
We measure the CPU performance on a Dell PowerEdge R750 server with two Intel Xeon Gold 6346 CPUs. 
We modify the GMPbench 6.2.1 to enable multi-thread execution.
We measure the single-core performance and also the 32-thread, and 64-thread performance on the CPU server.
We choose 32-thread performance as it is higher than 64-thread and calculate the energy efficiency by dividing the total power of two CPU cores, i.e., 205 Watts x 2 = 410 Watts.
For GPU measurement, we adopt perf\_tests provided in GPU CGBN~\cite{gpu_cgbn} library, and the power consumption is measured using nvidia-smi~\cite{nvidia-smi}. 

\vspace{-5pt}
\subsection{AIM On-board Implementation Results and Discussions}
\begin{figure}
\vspace{-5pt}
\centering
\includegraphics[width=0.7\linewidth]{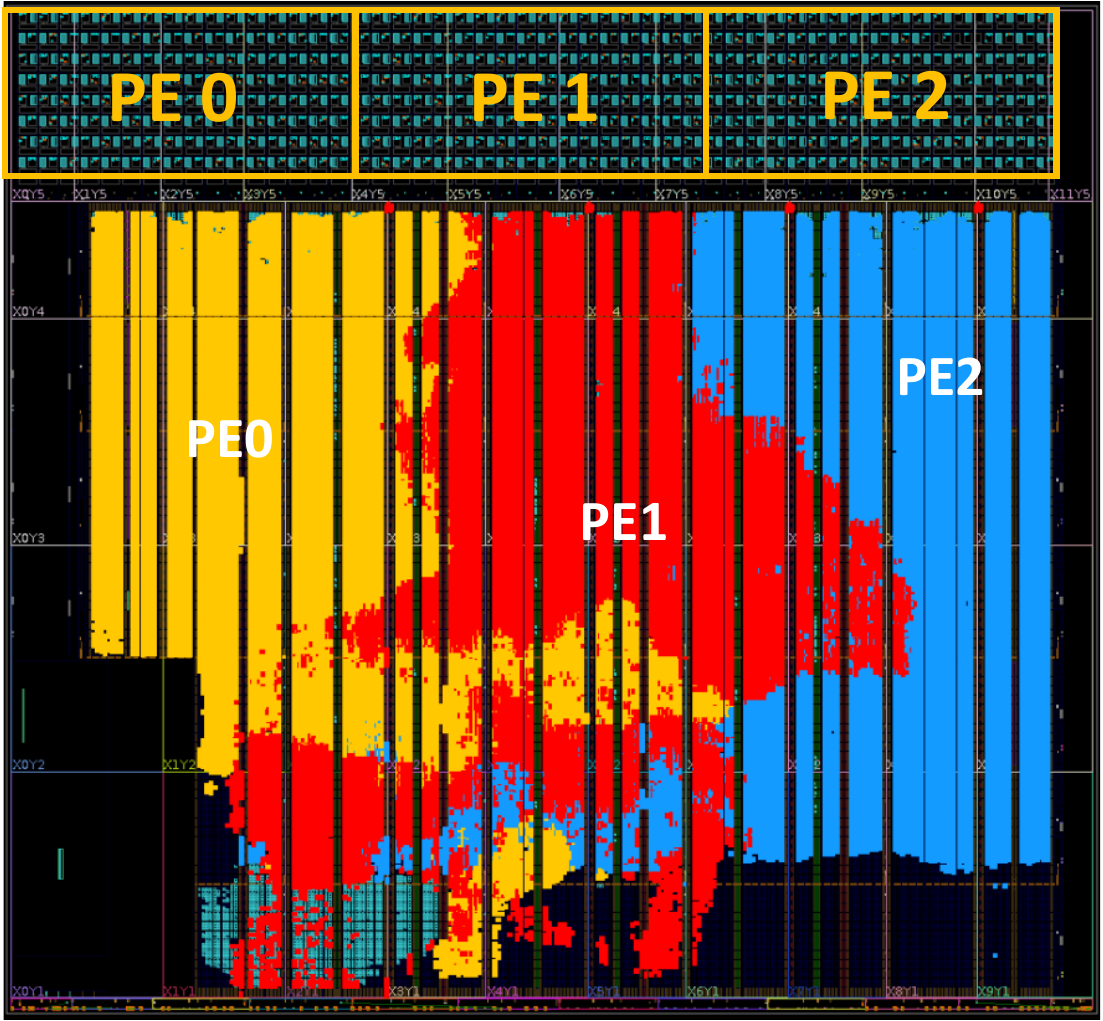}
\vspace{-5pt}
\caption{Layout of the optimal design point for 65,536-bit LIM.}
\label{fig:64k_layout}
\end{figure}

\vspace{-3pt}
\noindent\textbf{Model Accuracy.}
To verify the accuracy of the analytical model, we select different configurations of inter-task parallelism and intra-task parallelism for 65,536-bit LIM. 
Table~\ref{tbl:dse_accuracy} shows the comparison between the analytical models and the on-board measurement. 
The max error rate is 6.9\% and the average error rate is 1.8\%, which shows that our analytical models achieve good accuracy. 
The maximal throughput can be achieved with inter-task parallelism equal to 3 and intra-task parallelism equal to 132. This configuration uses 396 AIEs, and the implementation layout is shown in Figure~\ref{fig:64k_layout}.

\noindent\textbf{Comparisons among AIM, FPGA, CPU, GPU, and ASIC.}
We leverage AIM-DSE to search for optimal configurations for application LIM with data sizes from 4,096-bit to 262,144-bit. Table~\ref{tbl:single_kernel_comparison} compares performance, and energy efficiency among Versal AIM, CPU GMP, and GPU CGBN respectively. 
AIM Architecture achieves up to 1.43x throughput gain over the CPU server which has two Intel 10nm Xeon 6346 CPUs, in total, 32-cores. 
AIM achieves 
44.61x
throughput gain over a single CPU thread.  It is worth mentioning that the CPU GMP baseline does not only adopt schoolbook decomposition. 
We use Intel Vtune~\cite{vtune} to obtain the function call stack and find that more advanced decomposition methods such as toom-cook~\cite{cook1969minimum}, etc. are adopted, which reduce the number of required multiplications by introducing more additions and memory footprints. 
This is the reason that the throughput gap between CPU GMP and AIM drops. 

In terms of energy efficiency, AIM achieves up to 12.6x and 2.1x gains over Intel Ice Lake 6346 CPU and Nvidia A5000 GPU. 
Note that AIM achieves similar or better performance with less than 77 watts of total power in contrast to 410 watts of CPUs and 230 watts of the GPU A5000. 
Compared to the FPGA IMpress~\cite{fccm_impress} accelerator on Alveo U250, AIM achieves up to 46.7x energy efficiency gain. We believe that we can achieve  higher performance for AIM if we combine different decomposition methods adopted in CPU GMP, and we leave this as our future work.
Compared to ASIC design Cambricon-P~\cite{cambricon_p}, AIM achieves 2.21x throughput gain.
Indeed AIM consumes 13x more power than Cambricon-P.
However, to be noted, we achieve ASIC-like performance by designing accelerators on a reconfigurable computing platform with a cost of \$10K in contrast to designing a customized 14nm ASIC chip which costs over \$100M~\cite{chipCostMckinsey}.

\input{table/RSA_Comparison}
\noindent\textbf{RSA Encryption.}
We compare AIM in accelerating more complex applications, e.g., RSA, with CPU using GMPBench~\cite{cpu_gmp} library in Table~\ref{tbl:rsa_comparison}. 
We do not include GPU results because the GPU CGBN library does not provide an RSA implementation. 
AIM achieves up to 380x throughput gain and 1966.6x energy efficiency gain over Intel Xeon Gold 6346. 
We look into CPU GMPBench and find that CPU RSA adopts a different algorithm and spends a lot of time computing large integer modulo operations. 
AIM adopts an alternative efficient algorithm that transforms modulo operations into shift and multiplication. 
Still, in AIM RSA implementation, the AIE kernels are fully pipelined, and the latency of the other kernels that are implemented on PL is hidden, which does not introduce extra execution time in the end-to-end applications.  
This is different from the CPU programming model where non-accelerated kernels easily diminish the performance gain from the accelerated kernels on AVX instructions.

\noindent\textbf{Mandelbrot Set.}
\input{table/Mandelbrot_Comaprison}
Table~\ref{tbl:mandelbrot_comparison} shows the comparisons among AIM, Intel Xeon 6346 CPU (GMP 6.2.1), and Nvidia A5000 GPU (CGBN 2.0) in plotting the same area of the Mandelbrot set using the same configuration. 
We use a single CPU core as the baseline. 
AIM achieves up to 8.62x and 1.73x energy efficiency gains over CPU and GPU respectively. 
 The energy efficiency gains are smaller than that in LIM (Table~\ref{tbl:single_kernel_comparison}). 
One reason is that Mandelbrot heavily computes in square multiplication and the CPU GMP library calculates faster when two operands are the same than when calculating two different operands.
We leave this optimization in AIM in our future work.

%% file: table/single_kernel_comparison.tex
\begin{table*}[!tb]
\footnotesize
\caption{Optimal AIM Implementation for LIM with input sizes from 4,096-bit to 262,144-bit. We show performance and energy efficiency comparisons among AIM, Intel 10nm Xeon 6346 CPU, and Nvidia A5000 GPU. For GPU, \textcolor{red}{$\times$} means it is not supported in the library.}
\vspace{-5pt}
\label{tbl:single_kernel_comparison}
\begin{adjustbox}{width=2\columnwidth,center}
\begin{tabular}{c | c c | c c | c c | c c }
 \toprule
& \multicolumn{2}{c|}{\textbf{CPU (32 cores, 410W)}}
& \multicolumn{2}{c|}{\textbf{GPU (230W)}}
& \multicolumn{2}{c|}{\textbf{AIM ($<$77W)}}
& \multicolumn{2}{c}{\textbf{Energy Eff. Gain}} \\
   \textbf{Input Bits}  
  &  \textbf{kTasks/s} &  \textbf{kTasks/s/Watt}  &  \textbf{kTasks/s} &  \textbf{kTasks/s/Watt}
  &  \textbf{kTasks/s} &  \textbf{kTasks/s/Watt}  &  \textbf{AIM vs CPU} &\textbf{AIM vs GPU}\\
    \midrule
   4,096   & 23,259 & 56.73  & 145,474 & 632.50                                         & 17,685  & 467.87  & 8.25x  & 0.74x \\
   8,192   &  7,619 & 18.58  & 36,760  & 159.83                                         & 9,578   & 220.04  & 11.84x & 1.38x \\
   16,384  &  2,726 & 6.65   & 11,355  & 49.37                                          & 3,901   & 84.02   & 12.63x & 1.70x \\
   32,768  &  1,026 & 2.50   & 2,970   & 12.91                                          & 1,438   & 27.46   & 10.96x & 2.13x \\
   65,536  &   386.0 & 0.94   & \textcolor{red}{$\times$}  & \textcolor{red}{$\times$}     & 459.8    & 6.86    & 7.29x  & \textcolor{red}{$\times$} \\
   131,072 &   145.3 & 0.35   & \textcolor{red}{$\times$}  & \textcolor{red}{$\times$}     & 128.1    & 1.75    & 4.93x  & \textcolor{red}{$\times$} \\
   262,144 &    57.0 & 0.14   & \textcolor{red}{$\times$}  & \textcolor{red}{$\times$}     & 33.8     & 0.44    & 3.15x  & \textcolor{red}{$\times$} \\

    \bottomrule
\end{tabular}
\end{adjustbox}
\vspace{-15pt}
\end{table*}

%% file: table/RSA_Comparison.tex
\begin{table}
\footnotesize
\caption{Performance and energy efficiency comparison between GMP on Intel 10nm Xeon 6346 CPU (32 core, 410 Watt) and AIM on VCK190 for RSA.}
\vspace{-5pt}
\label{tbl:rsa_comparison}
\begin{adjustbox}{width=0.9\columnwidth,center}
\begin{tabular}{c | c c | c c }
 \toprule
& \multicolumn{2}{c|}{\textbf{CPU}}
& \multicolumn{2}{c}{\textbf{AIM}} \\
   \textbf{Input Bits}  
  &  \textbf{Tasks/s} &  \textbf{Tasks/s/Watt}  &  \textbf{Tasks/s} &  \textbf{Tasks/s/Watt} \\
    \midrule
   4,096    & 6124  & 14.97 (1x) & 81734 & 2458.2 (162.6x)  \\
   8,192    & 930   & 2.27 (1x) & 44737 & 1196.2 (527.2x)  \\
   16,384   & 161   & 0.39 (1x) & 19017 & 435.2 (1109.2x)  \\
   32,768   & 28    & 0.07 (1x) & 10639 & 134.8 (1966.6x)  \\
    \bottomrule
\end{tabular}
\end{adjustbox}
\end{table}

%% file: table/Mandelbrot_Comaprison.tex
\begin{table}
\footnotesize
\caption{Performance and energy efficiency comparisons among GMP~\cite{cpu_gmp} on Intel 10nm Xeon 6346 CPU, and CGBN~\cite{gpu_cgbn} on Nvidia 8 nm A5000 GPU (230 Watt), and AIM (ours) on VCK190 for plotting Mandelbrot set.}
\vspace{-5pt}
\label{tbl:mandelbrot_comparison}
\begin{adjustbox}{width=1\columnwidth,center}
\begin{tabular}{c | c c | c c | c c }
 \toprule
& \multicolumn{2}{c|}{\textbf{CPU}}
& \multicolumn{2}{c|}{\textbf{GPU}}
& \multicolumn{2}{c}{\textbf{AIM}} \\
   \textbf{Input Bits}  
  &  \textbf{Tasks/s} &  \textbf{Tasks/s/Watt}  &  \textbf{Tasks/s} &  \textbf{Tasks/s/Watt} & \textbf{Tasks/s} &  \textbf{Tasks/s/Watt}\\
    \midrule
   8,192    & 0.048   & 0.0037 (1x) & 6.790 & 0.0326 (8.80x) & 0.641 & 0.0228 (6.15x)\\
   16,384   & 0.016   & 0.0013 (1x) & 1.799 & 0.0087 (6.74x) & 0.241 & 0.0088 (6.85x)\\
   32,768   & 0.006   & 0.0005 (1x) & 0.509 & 0.0024 (4.99x) & 0.126 & 0.0042 (8.62x)\\
    \bottomrule
\end{tabular}
\end{adjustbox}
\vspace{-3pt}
\end{table}

%% file: 7_conclusion.tex
\vspace{-10pt}
\section{Conclusion}
\vspace{-5pt}
In this work, we first analyze the energy efficiency gains when mapping arbitrary-precision integer multiplication onto CPUs and GPUs over reconfigurable computing, e.g., FPGA comes from the vector units in CPUs and GPUs. 
We propose the AIM, a customized accelerator architecture on Versal ACAP, i.e., a new heterogeneous reconfigurable computing platform with added vector processors. 
We propose the AIM framework that can systematically generate and optimize AIM designs. 
We integrate AIM architecture into multiple end-to-end applications and demonstrate that AIM achieves the highest energy efficiency among the SOTA accelerators and libraries including CPUs, GPUs, and FPGA. 
We will explore the other decomposition methods and more applications in future work. 
\vspace{-4pt}
\section*{Acknowledgements}
\vspace{-2pt}
We acknowledge the support from the University of Pittsburgh New Faculty Start-up Grant, 
National Science Foundation (NSF) awards \#1822085, \#2019336, \#2213701, \#2217003, \#2229562, the Laboratory of Physical Sciences (LPS), and NSF's Cloud Access program CloudBank.
We thank all the reviewers for their valuable feedback. 
We thank AMD/Xilinx for hardware and software donations.
\vspace{-10pt}